\documentclass[sigconf]{acmart}
\usepackage{amsmath}
\usepackage{booktabs}
\usepackage{soul}
\usepackage{url}
\usepackage{scalerel}
\usepackage{balance}

\usepackage[ruled,vlined]{algorithm2e}
\usepackage{algpseudocode}

\author{Zheng Liu}
\authornote{These authors contribute equally to this work.}
\affiliation{%
  \institution{Microsoft Research Asia}
  \city{Beijing}
  \state{China}
}
\email{zheng.liu@microsoft.com}

\author{Yu Xing}
\authornotemark[1]
\affiliation{%
  \institution{USTC}
  \city{Los Angeles}
  \state{CA, USA}
}
\email{xingy@usc.edu}

\author{Jianxun Lian}
\authornotemark[1]
\affiliation{%
  \institution{Microsoft Research Asia}
  \city{Beijing}
  \state{China}
}
\email{jialia@microsoft.com}

\author{Defu Lian}
\affiliation{%
  \institution{USTC}
  \city{Hefei}
  \state{China}
}
\email{liandefu@ustc.edu.cn}

\author{Xing Xie}
\affiliation{%
  \institution{Microsoft Research Asia}
  \city{Beijing}
  \state{China}
}
\email{xingxie@microsoft.com}

\author{Ziyao Li}
\affiliation{%
  \institution{Carnegie Mellon University}
  \city{Pittsburg}
  \state{PA, USA}
}
\email{ziyaoli@andrew.cmu.edu}

\begin{document}

\title{A Novel User Representation Paradigm for Candidate Retrieval}


\begin{abstract}
    Candidate retrieval is a fundamental issue in recommendation system. Given user's recommendation request, relevant candidates need to be retrieved in realtime for subsequent ranking operations. Considering that the retrieval operation is conducted over considerable items, it has to be both precise and scalable so that high-quality candidates can be acquired within tolerable latency. Unfortunately, conventional methods would trade off precision for high running efficiency, which leads to inferior retrieval quality. In contrast, those deep learning-based approaches can be highly accurate in identifying relevant items; yet, they are unsuitable for candidate retrieval due to their inherent limitation on scalability.

    In this work, a novel framework is proposed to address the above challenges. The underlying intuition is to rely on a well-trained ranking model for the supervision of an efficient retrieval model, such that it will unify the scalability and precision as a whole. We have implemented our conceptual framework and made comprehensive evaluation for it, where promising results are achieved against representative baselines.
    Our work is undergoing a anonymous review, and it will soon be released after the notification. If you're also interested in this problem, please feel free to contact us.
\end{abstract}


\maketitle

\section{Introduction}\label{sec:intro}
\begin{figure}[t]
\centering
\includegraphics[width=\linewidth]{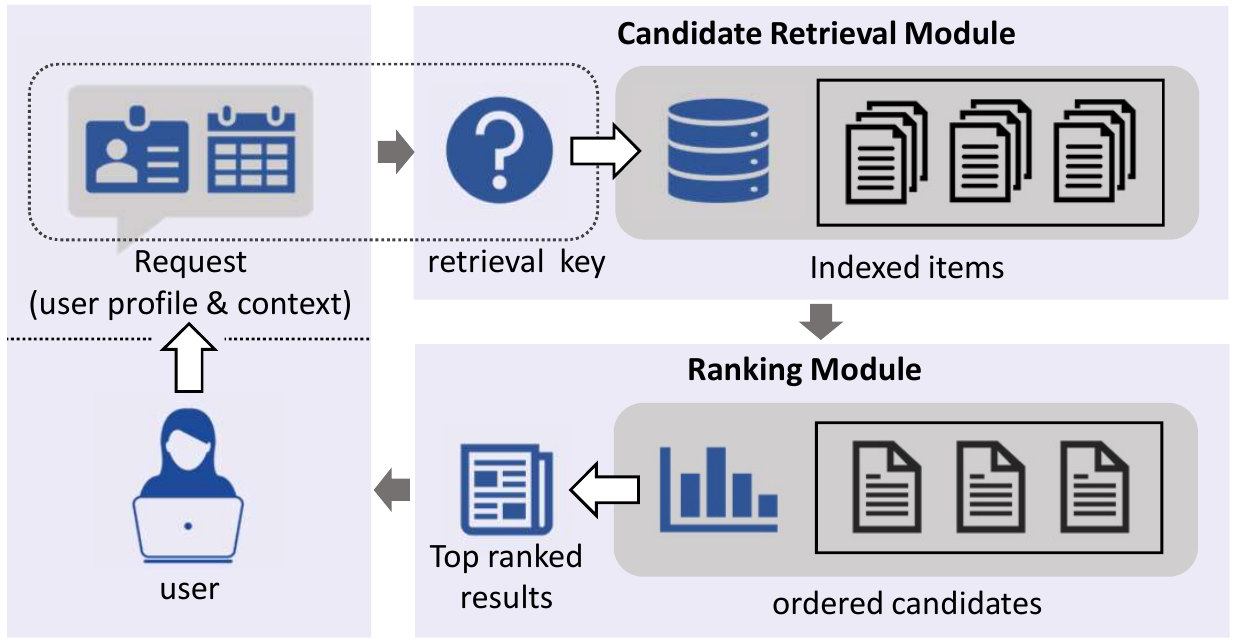}
\caption{Sketch of the typical workflow for a recommendation system (better viewed in colour).}
\label{fig:workflow}
\end{figure}

Recommendation system plays a crucial role in modern web services, e.g., online advertisement and e-commerce, as users' interested items can be automatically delivered on top of the analysis of their intensive behavioral data. Instead of selecting suitable items in-one-shot, a typical recommendation system would consecutively execute two fundamental operations (illustrated as Figure \ref{fig:workflow}): \textbf{candidate retrieval} and \textbf{ranking} \cite{cheng2016wide,covington2016deep}. Given a recommendation request from user, the retrieval module will select a small set of relevant candidates in realtime from the tremendous pool of items; then, the ranking module will further refine the candidates with higher precision, where those top ranked candidates are to be generated as the final recommendation result. It is apparent that the candidate retrieval operation severely affects the overall recommendation quality, whose performance has to be optimized within the integral system.

Because of its distinct role in the recommendation system, candidate retrieval is desirable of satisfying two pieces of properties. For one thing, considering tremendous the scale of item set in reality, candidate retrieval must be temporally scalable so as to maintain a tolerable running cost. For anther thing, to have high quality retrieval result, user interest must be precisely modeled while searching for appropriate candidates.

Conventionally, it is a common practice \cite{ChenW18} to represent user and item with a shared group of features (e.g., keywords, categorical tags or LDA vectors), so that relevant candidates of a user can be retrieved based on their feature similarity. During such an operation, index structures like LSH (locality sensitive hashing) and KNN graph are pre-constructed for all the items, which will significantly short-cut the retrieval process thanks to their superior efficiency on similarity search. Following the same candidate retrieval paradigm, more proficient algorithms are developed on top of metric learning, such as DSSM \cite{huang2013learning} and CML \cite{hsieh2017collaborative}, where more expressive latent features are learned for user and item to better characterize their mutual relevance. Regardless of different formulations, all the above approaches have to represent user and item in the same space, and retrieve the candidates purely based on feature similarity (e.g., Cosine or Euclidean distance between user and item's feature vectors). Such a highly restricted workflow will probably give rise to inferior modeling of user interest, thus harming the retrieval quality.

By comparison, various types of deep learning-based recommenders have been developed recently, such as Wide\&Deep \cite{cheng2016wide}, Deep\&Cross \cite{wang2017deep}, DeepFM \cite{guo2018deepfm} and xDeepFM \cite{lian2018xdeepfm}, where user interest can be captured with high precision. However, such approaches would rely on complex functional relationship between user and item's features; in other words, user interest is no longer reflected by feature similarity.
As a result, it is hard to index the items w.r.t. the user interest learned by such recommenders. Therefore, they are mainly used for the ranking operation rather than candidate retrieval.

In short, conventional methods based on feature similarity are scalable for realtime retrieval, but their retrieval quality can be limited by inferior user interest modeling. Meanwhile, deep learning-based recommenders are precise in modeling user interest, yet they are difficult to be scaled for candidate retrieval.

In this work, we propose a novel personalized candidate retrieval paradigm Recallnet, which makes the best of both worlds. On the one hand, user interest is precisely captured by an arbitrary type of deep recommendation model; on the other hand, relevant candidates can be directly obtained via similarity search, which enables items to be indexed for efficient retrieval. The underlying idea of Recallnet is quite intuitive. Particularly, a deep recommendation $\mathbf{D}$ (e.g., \cite{cheng2016wide,wang2017deep,guo2018deepfm,lian2018xdeepfm}) is employed for the precise modeling of user interest; 
meanwhile, another model $\mathbf{\Psi}$ is learned to synthesize a ``virtual item representation'' for a user, referred as the \textbf{retrieval key}. Importantly, it is expected that user interest captured by $\mathbf{D}$ can be well approximated by the similarity between retrieval key and those real items' representations. In other words, given user $u$, who prefers item $a$ over $b$; then u's retrieval key will always be more similar with $a$'s representation:
\begin{align}\label{eq:rule}
\begin{split}
\mathrm{sim}(\mathbf{\Psi}_u, \theta_a) &> \mathrm{sim}(\mathbf{\Psi}_u, \theta_b) \\
\text{if } \mathbf{D}(u, a) &> \mathbf{D}(u, b),
\end{split}    
\end{align}
where $\mathrm{sim}(\cdot)$ is a certain similarity function, e.g., Cosine similarity, $\mathbf{\Psi}_u$ indicates the retrieval key, and $\theta$ stands for the item's representation. 
Given the satisfaction of the above relationship, user's interested items will be confined within the neighborhood of $\mathbf{\Psi}_u$; thus, it will enable high relevance candidates to be retrieved efficiently via similarity search. 

To realize the above candidate retrieval paradigm, an Actor-Critic style learning framework \cite{Konda00actor-criticalgorithms,bhatnagar2009natural} is developed. With the Actor module, the retrieval key is synthesized for each user, together with representations generated for each item, based on both parties' inherent information. And with the Critic module, supervision signals are generated for Actor, such that the relationship in Eq. \ref{eq:rule} can be achieved. In Critic, a compound reward is integrated from three sources: \textbf{evaluator}, \textbf{validator}, and \textbf{referencer}, whose individual functionality is elaborated as follows. 

$\mathbf{(1)}$ The evaluator is an arbitrary form of deep recommendation model, which is pre-trained for the accurate modeling of user's underlying interest. Now that the retrieval key can be regarded as the representation of a virtual item, the well-trained evaluator is employed to measure user's degree of interest towards such a virtual item. 

$\mathbf{(2)}$ Because of deep models' inherent unrobustness \cite{goodfellow2014explaining,carlini2017towards,madry2017towards}, the evaluator might falsely reward an inferior synthesization. To get rid of this potential defect, an auxiliary discriminative model called validitor is introduced, which resists the evaluator from being fooled by adversarial cases.

$\mathbf{(3)}$ The referencer is a certain type of similarity measurement, e.g., Cosine or Euclidean, which is used to encourage the maximization of similarity between the retrieval key and candidates' representations.

\begin{figure}[t]
\centering
\includegraphics[width=0.8\linewidth]{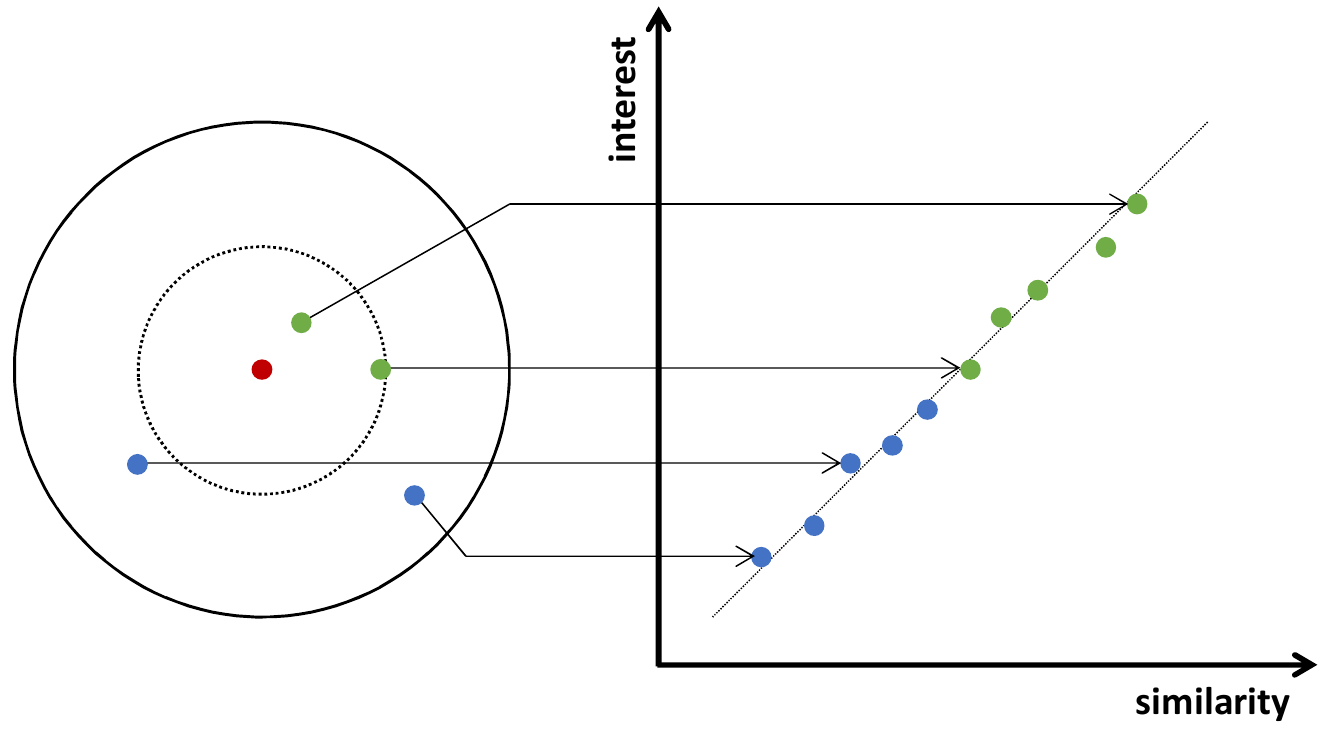}
\caption{Illustration of Recallnet's objective: to have the similarity between the retrieval key and item's representation aligned with user interest.}
\label{fig:eg-intro}
\end{figure}

With the maximization of rewards in (1) and (2), $\mathbf{\Psi}_u$ will become a local optima of user interest; in addition, with the maximization of reward in (3), representations of user's interested items will be anchored in the neighborhood of $\mathbf{\Psi}_u$. Finally, by maximizing all the rewards simultaneously, the ``concentric diagram'' is formed (shown as Figure \ref{fig:eg-intro}), with user's most interested point $\mathbf{\Psi}_u$ being the center (the red point), representations of user's interested items being $\mathbf{\Psi}_u$'s neighborhood (the green points), and representations of those less interested items distributed away from $\mathbf{\Psi}_u$ (the blue points). In other words, user interest becomes almost positively related with the similarity between retrieval key and item representation. Therefore, the relationship stated by Eq. \ref{eq:rule} will come into existence. 

To summarize, the major contribution of this work is highlighted as follows.
\begin{enumerate}
    \item A novel paradigm Recallnet is proposed in this work, which learns to synthesize the virtual item representation (i.e., retrieval key) for both precise and temporally scalable candidate retrieval.
    
    \item An Actor-Critic infrastructure is developed for Recallnet, which enables user interest to be well approximated by the similarity between retrieval key and item representation.
    
    \item Extensive empirical studies are conducted with a series of real-world datasets, where consistent and remarkable improvements are achieved representative baseline methods, such as those based on metric-learning.
\end{enumerate}

The subsequent contents of this work are organized as follows. First of all, preliminaries and formulation of our problem are presented in Section \ref{sec:def}. Secondly, Recallnet's architecture is over-viewed in Section \ref{sec:sys}, followed by its instantiation discussed in Section \ref{sec:ins}. The experimental studies are made in Section \ref{sec:exp}; and the related works are reviewed in Section \ref{sec:rel}. Finally, the paper is concluded in Section \ref{sec:con}.


\section{Problem Formulation}\label{sec:def}
In this section, preliminaries for recommendation system and candidate retrieval are introduced in the first place, on top of which definition and quantitive formulation are presented for personalized candidate retrieval problem. 

\subsection{Preliminaries and Definition}
\subsubsection{Typical Workflow of Recommendation} 
As sketched by Figure \ref{fig:workflow}, a typical recommendation system operates with the following two consecutive steps.

$\bullet$ Candidate Retrieval. Given a recommendation request, which consists of user and context information, a small group of relevant candidates are retrieved in realtime from the whole item set.

$\bullet$ Candidate Ranking. The retrieved candidates will be further refined by the ranking module, which is of higher precision; the top ranked candidates will be returned as the recommendation result.

Given both parts' distinction in functionality, different techniques have to be developed in practice. Particularly, the ranking module mainly emphasizes on accuracy, whose developed algorithm should be as precise as possible; in contrast, the candidate retrieval module has to jointly consider precision and scalability, given that the whole items' scale could be huge in reality.

\subsubsection{Candidate Retrieval} 
Despite diverse formulations in detail, mainstream candidate retrieval methods share the common framework. Particularly, they would first determine the way of representation for the items (e.g., bag-of-keywords), along with their similarity measurement (e.g., keyword co-occurrence). Then items are organized with a certain index structure (e.g., hashing tables), where similar items can be grouped in common units. Once a recommendation request is issued from a user, the retrieval key will be generated for her, which follows the identical representation format as that of item. Finally, the retrieval key is used to search the pre-constructed index, where items with similar representations will be retrieved as candidates. Thanks to the sub-linear time complexity of such a similarity search paradigm, the candidate retrieval can be highly scalable, thus being able to be completed in realtime.

However, it remains an open question of designing better schemes so that high-relevance candidates can be comprehensively acquired. 
Particularly, the similarity search paradigm must be well aligned with user's underlying interest, such that user's top interested items will also be those whose representations are highly similar with the retrieval key.
With the above concern in mind, the optimal personalized candidate retrieval (OPCR) problem is defined as follows.
\begin{definition}\label{def:1}
In OPCR, user's most interested items will also become those whose representations are most similar with the retrieval key; therefore, the top relevance candidates can be retrieved efficiently via similarity search.
\end{definition}

\subsection{Quantative Formulation}
In this part, the quantitive formulation of OPCR is presented, which also helps to illustrate our intuition of solving it in practice.

Let $u$/$v$ be an arbitrary user/item, respectively. Suppose there is an ``almost perfect'' deep recommendation model $\mathbf{D}$, which precisely measures user interest as $\mathbf{D}(\theta_u,\theta_v)$, where $\theta_{*}$ stands for the representations of the corresponding entity. In addition, there is a synthesization model $\mathbf{\Psi}$, which generates $\mathbf{\Psi}_u$ as the retrieval key.
With a recommendation request $r$ from user $u$, the following optimization problem is formulated, which specifies the retrieval of user's most interested candidates:
\begin{align}\label{eq:opt}
\begin{split}
    \max_{\mathbf{\Psi}, \theta_{*}} \sum\nolimits_{\Omega_r} \mathbf{D}(\theta_u, \theta_v)& \\
    s.t. ~~\textbf{ 1) } \theta_v \in \text{KNN}(\mathbf{\Psi}_u),  ~ \forall v \in \Omega_r;& ~~\textbf{ 2) } \Lambda_r \subset \Omega_r;
\end{split}
\end{align}
where $\Omega_r$ stands for the retrieved candidates. The first constraint indicates that every candidate's representation is among the top-K nearest neighbours to $\mathbf{\Psi}_u$. Meanwhile, the second constraint requires the incorporation of the ``ground truth'', i.e., user's consumed items in reality (whose validity as candidates is self-evident) needs to be included in the selected candidates. 

As is formulated by Eq. \ref{eq:opt}, the OPCR looks for the optimal configuration of synthesization model $\mathbf{\Psi}$ and item representation $\theta_{*}$, which will give rise to the retrieval of the best candidates. However, because of the combinatorial nature of Eq. \ref{eq:opt}, it will probably be difficult to obtain the optimal solution. As a result, a few mild relaxations are introduced for its approximate solution. 

Firstly, we would let the KNN requirement replaced by a threshold constraint on similarity:
\begin{align}\label{eq:relaxed-1}
\begin{split}
\max_{\mathbf{\Psi}, \theta_{*}} \sum\nolimits_{\Omega_r} \mathbf{D}(\theta_u,\theta_v) & \\
s.t. ~~\textbf{ 1) } \mathrm{sim}(\mathbf{\Psi}_u, \theta_v) \geq \epsilon, ~  \forall v \in \Omega_r; & ~~\textbf{ 2) } \Lambda_r \subset \Omega_r.
\end{split}
\end{align}
where $\mathrm{sim}(\cdot)$ indicates the similarity measurement, and $\epsilon$ stands for the similarity threshold which filters high-relevance candidates from the whole items. Since $\mathbf{\Psi}_u$ and $\theta_v$ are required to be highly similar, we may adapt the objective function by approximation:
\begin{align}\label{eq:relaxed-2}
\begin{split}
& \max_{\mathbf{\Psi}, \theta_{*}} \mathbf{D}(\theta_u,\mathbf{\Psi}_u) \\
s.t. ~~\textbf{ 1) } \mathrm{sim}(\mathbf{\Psi}_u, & \theta_v) \geq \epsilon, ~  \forall v \in \Omega_r; ~~\textbf{ 2) } \Lambda_r \subset \Omega_r.
\end{split}
\end{align}
Intuitively, it requires the maximization of user's interest towards $\mathbf{\Psi}_u$; thus, user's interest to the candidates can be maximized as well, thanks to their representations' high similarity with $\mathbf{\Psi}_u$. Moreover, the similarity constraint is relaxed from $\Omega_r$ to $\Lambda_r$, and the above problem is transformed into its Lagrangian relaxation form:
\begin{equation}\label{eq:lag}
    \max_{\mathbf{\Psi}, \theta_{*}} \mathbf{D}(\theta_u,\mathbf{\Psi}_u) + \lambda * \sum\nolimits_{\Lambda_r} \mathrm{sim}(\mathbf{\Psi}_u, \theta_v).
\end{equation}
Finally, it it constraint-free and fully differentiable (given that common similarity measurement, e.g., $\mathrm{cos}(\cdot)$ or $\|\cdot\|_2$, are adopted). Therefore, it becomes solvable via gradient ascent.

Now we may come to the following high-level framework for OPCR's solution. 

$\bullet$ Firstly, a certain type of deep recommendation model $\mathbf{D}$ is selected and pre-trained based on user history, where user interest can be accurately predicted;

$\bullet$ Secondly, the synthesization model $\mathbf{\Psi}$ and item representation $\theta_{*}$ are trained to maximize the objective function in Eq. \ref{eq:lag} via gradient ascent. In this place, both functions $\mathbf{D}(\cdot)$ and $\mathrm{sim}(\cdot)$ are employed as the ``critics'' for $\mathbf{\Psi}$ and $\theta_{*}$'s performances, whereby providing the supervision signals for both parties' iterative updates.

Finally, we will get the near optimal solution to the original optimization problem in Eq. \ref{eq:opt}, which realizes our proposed objectives in OPCR.


\section{General Infrastructure of Recallnet}\label{sec:sys}
Recallnet is a abstractive candidate retrieval paradigm, whose concrete realization can be flexibly adapted according to specific application scenarios. In this section, an overview is made for Recallnet's general infrastructure, which is illustrated with Figure \ref{fig:frame}. Our discussion is partitioned into two parts: 1) Recallnet's offline construction and 2) its online workflow.

\subsection{Recallnet's Construction} 
Recallnet is constructed in the offline stage. The construction is carried out in an Actor-Critic pipeline shown as Figure \ref{fig:frame} (I). 

\subsubsection{Actor's Role} The Actor is to generate synthesized retrieval key and item representation based on user and item's raw features. Particularly, the Actor includes three components: (1) the synthesization module, where user's recommendation request is encoded as the retrieval key; (2) the user encoder and (3) the item encoder, where latent vectors are generated for user and item as their representations. It's worth nothing that the retrieval key and user representation are generated for different purposes: one for candidate retrieval, thus needs to follow the same format as the item representation; while the other one is for the deep recommendation model, whose format can be chosen flexibly for the best performance. 

\subsubsection{Critic's Role} The Critic is to provide supervision signals for the Actor's generation process. Particularly, a compound reward is integrated from three sources: evaluator, validator, and referencer. 

$\bullet$ The evaluator is a pre-trained deep recommendation model, which determines user's interest towards an item given both parties' representations. 
Once deployed in the Critic, it is used to calculate user's degree of interest towards the retrieval key (i.e., the representation of a virtual item). Such a reward can be interpreted as $\mathbf{D}(\cdot)$ in Eq. \ref{eq:lag}.

$\bullet$ The validator is a discriminative model, which tells whether the synthesized retrieval key is within the distributed scope of those real items' representations. Such a reward helps to eliminate the adversarial cases, where the evaluator will falsely reward a meaningless synthesization\footnote{The evaluator will become ineffective and generate unreliable reward, once the retrieval key is out of its working domain, which is also the distributed scope of real items' representations}.

$\bullet$ The referencer compares the similarity between the retrieval key and the representation of a candidate item. Such a reward is corresponding to $\mathrm{sim}(\cdot)$ in Eq. \ref{eq:lag}, which encourages the retrieval key and its candidates to be located in the same neighborhood.

\subsubsection{Construction Process} Given the above framework, Recallnet's construction is carried out via the interaction between Actor and Critic. First of all, the evaluator and validator are learned so that the Critic can be deployed before hand. Secondly, given user's historical behaviors, retrieval keys and user/item representations are consecutively generated by the Actor, which will further get rewarded by the Critic. The Actor will then be updated ascendingly w.r.t. the partial gradients of its acquired reward. 
Finally, the above generation-reward-update process will be iteratively conducted until its convergence, where the compound reward can be maximized.

\subsection{Recallnet's Workflow}
The Actor is isolated from the well constructed Recallnet and deployed for the candidate retrieval operation. 
Particularly, all the items are encoded as their representations and organized with a certain index structure in the offline stage. 
Once an recommendation request is issued by the online service, the retrieval key will be synthesized and used for the ANN (approximate nearest neighbor) search over pre-constructed index. 
Finally, the top-K similar items can be identified, which will be returned as the retrieved candidates.


\section{Recallnet's Instantiation}\label{sec:ins}
As introduced, Recallnet is a general paradigm for personalized candidate retrieval, whose formulation can be flexibly adapted for different recommendation scenarios. However, to better demonstrate how Recallnet works in practice, it is instantiated for ``text-rich'' scenario, which is common for a wide variety of applications, such as online advertisement, e-commerce and news recommendation (e.g., \cite{zhai2016deepintent,Tay2018multi,wu2019neural}). Particularly, settings for the discussed scenario are briefly introduced as follows.

$\bullet$ \textbf{Item.} Each item is associated with its text feature, which is organized as a sequence-of-word (e.g., the title of ad/news article). 

$\bullet$ \textbf{User.} Each user is associated with her historical behaviors, where each behavior stands for a specific item consumed by the user (e.g., the whole ads/news articles clicked in history). 

For the next part, concrete structures are designed so that Recallnet will be instantiated for the above scenario\footnote{Although there can be various alternative structures for Recallnet, only one representative is discussed here for demonstration. However, more necessary alternatives are to be analyzed in the empirical studies.}. To facilitate comprehension, the frequently used notations are summarized in Table \ref{tab:notation}.

\begin{table}[t]
\centering
\includegraphics[width=0.85\linewidth]{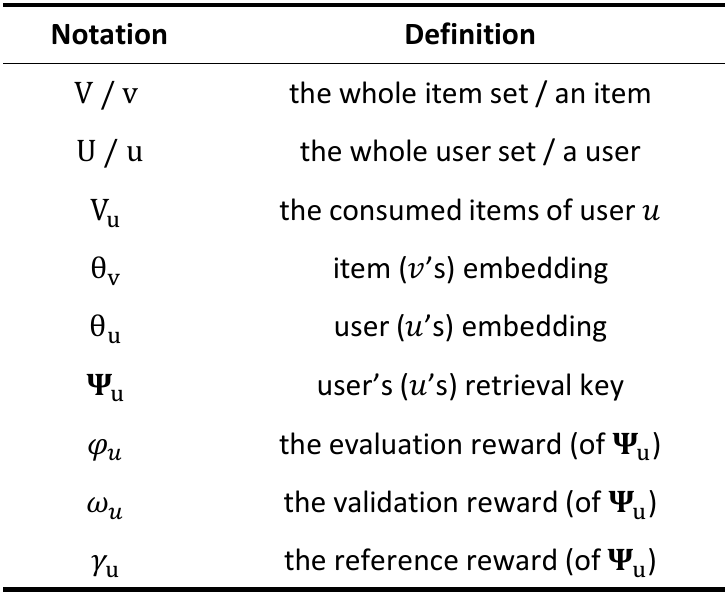}
\caption{Frequently Used Notations.}
\label{tab:notation}
\end{table}


\subsection{Actor}

\subsubsection{Item Encoder}
\begin{figure}[t]
\centering
\includegraphics[width=0.75\linewidth]{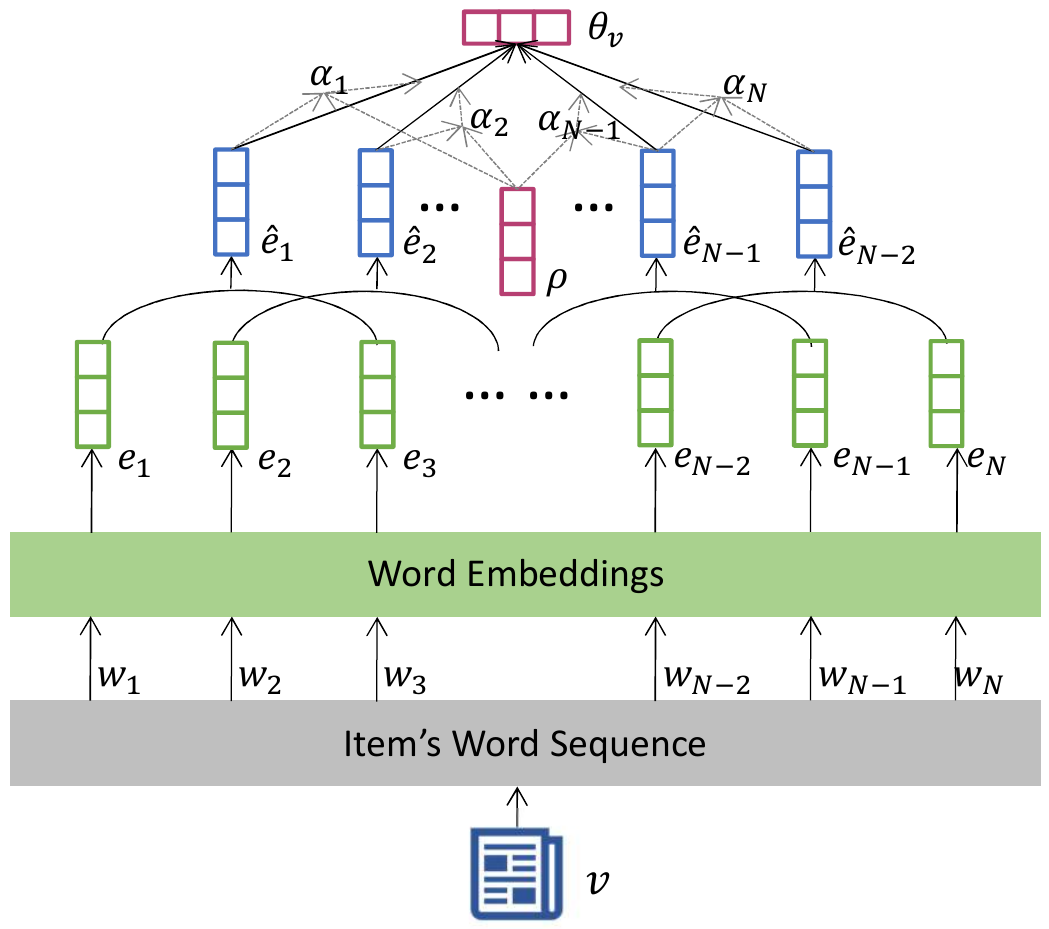}
\caption{Item encoder's structure.}
\label{fig:item}
\end{figure}

The item encoder's structure is shown by Figure \ref{fig:item}. First of all, the item's word sequence $\mathbf{w}_v$ is transformed into the word embedding sequence $\mathbf{e}_v$ by transferring each token into its embedding vector:
\begin{equation}\label{eq:w2v}
    \mathbf{e}_v: [e_1, ..., e_N]_v \leftarrow \mathbf{w}_v: [w_1, ..., w_N]_v.
\end{equation}
The word embedding sequence is further processed by a 1-D convolutional network (CNN) \cite{kim2014convolutional} so as to better extract its local information:
\begin{equation}\label{eq:cnn}
    \mathbf{\hat{e}}_v \leftarrow \mathrm{CNN}(\mathbf{e}_v).
\end{equation}
To highlight the meaningful information, the vector $\mathbf{\rho}_a$ is introduced, which attentively aggregates the whole sequence:
\begin{equation}\label{eq:at-cnn}
\mathbf{\theta}_v = \sum\nolimits_{i} \alpha_i \mathbf{\hat{e}}_i, ~\text{ where } \alpha_i = \frac { \mathrm{exp}(\mathbf{\hat{e}}^T_i \mathbf{\rho}_a)  } { \sum\nolimits_{j} \mathrm{exp}(\mathbf{\hat{e}}^T_j \mathbf{\rho}_a) }.
\end{equation}
Finally, the aggregated vector $\mathbf{\theta}_v$ is used as the item's representation.

\subsubsection{User Encoder}\label{sec:ins-user}
\begin{figure}[t]
\centering
\includegraphics[width=0.75\linewidth]{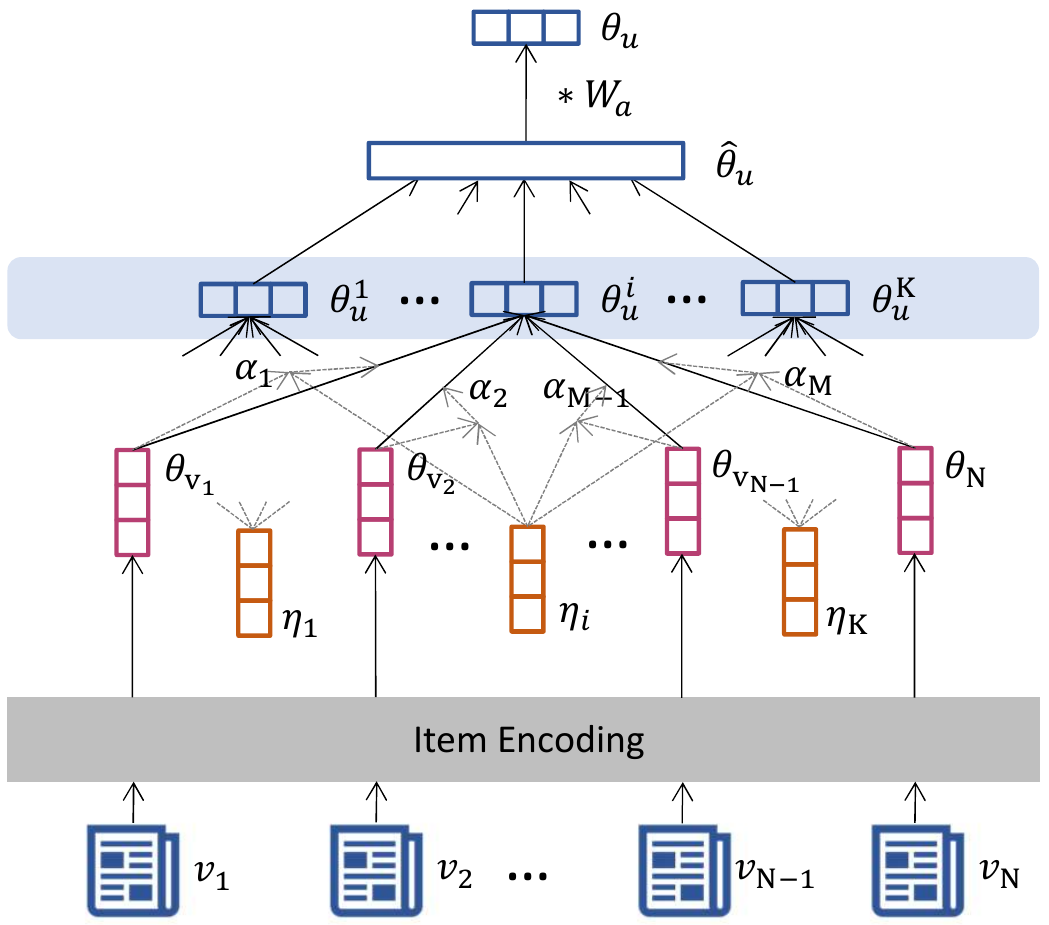}
\caption{User encoder's structure.}
\label{fig:user}
\end{figure}

The structure of user encoder is shown as Figure \ref{fig:user}. Particularly, user representation ($\mathbf{\theta}_u$) is generated by attentively aggregating the representations of user's consumed items $V_u$. In this place, multi-head attentive pooling \cite{lin2017structured} is employed for the aggregation. First of all, a total of K pooling heads $\{\mathbf{\eta_a^i}\}^{1,...,K}$ are employed, each of which is used to generate a unique aggregated vector via attention:
\begin{equation}\label{eq:at-user}
\theta_u^i = \sum\nolimits_{V_u} \alpha^i_v \mathbf{\theta}_v (i=1,...,K), ~\text{ where } \alpha^i_v = \frac { \mathrm{exp}(\theta_v^T \mathbf{\eta_a^i})  } { \sum\nolimits_{V_u} \mathrm{exp}(\theta_{v'}^T \mathbf{\eta_a^i}) }.
\end{equation}
As a result, a total of K aggregated vectors $\{\theta_u^i\}^{1...K}$ are obtained. All these vectors are concatenated along the column and multiplied by a $Kd\times d$ ($d$ is the dimension of $\theta_v$) mapping matrix $W_a$, where the original representation dimension will be kept:
\begin{equation}\label{eq:cat-user}
\theta_u = \mathrm{Relu}(W_a \bar{\theta}_u + b_a), ~\text{ where } \bar{\theta}_u = \mathrm{concatenate}(\{\theta_u^i\}^{1...K}).
\end{equation}

\subsubsection{Synthesizer}
The retrieval key ($\mathbf{\Psi}_u$) is synthesized via two consecutive steps: firstly, user representation $\theta_u$ is generated with the user encoder; secondly, a M-layer feed-forward network (FFN) is employed where user representation is translated into the retrieval key:
\begin{align}\label{eq:syn}
\begin{split}
\psi_u^{i+1} \leftarrow \mathrm{Relu} (W^i_g \psi_y^i & + b^i_g), ~~~ i = 0,...,M-1 ~ \text{and}~ \psi_u^0 = \theta_u; \\
\mathbf{\Psi}_u & \leftarrow W^M_g \psi_r^{M-1} + b^M_g;
\end{split}
\end{align}
where $W^i_g$ and $b^i_g$ stand for the mapping matrix and bias of the $i$-th perception layer. Notice that the activation function is removed for the last layer so that the retrieval key can be an arbitrary real value vector (otherwise it will be confined in certain scope and probably unable to approach its candidate's embedding).  


\subsection{Critic}

\subsubsection{Evaluator}\label{subsec:eva}
A bi-channel deep recommendation model is pre-trained for evaluation. Particularly, the item representation $\theta_v$ and user representation $\theta_u$ are delivered to two different multi-layer feed-forward networks: $\mathrm{FFN}_1$ and $\mathrm{FFN}_2$, where user's interest towards the item is measured with the weighted summation of both outputs' element-wise product:
\begin{align}\label{eq:rec}
\begin{split}
\varphi_{v,u} & = \log(  ~ \sigma(W_e (\hat{\theta}_v^T \odot \hat{\theta}_u)) ~), \\
\text{where}, ~ \hat{\theta}_v & \leftarrow \mathrm{FFN}_1( \theta_v), ~~ \hat{\theta}_u \leftarrow \mathrm{FFN}_2(\theta_u).
\end{split}
\end{align}
Here, $\sigma(\cdot)$ and $\odot$ indicate the sigmoid function and element-wise product, and $W_e$ is the learnable weighting vector. The well-trained recommendation model is deployed as the evaluator. It will treat the retrieval key $\mathbf{\Psi}_u$ as a virtual item's representation, and calculate the evaluation reward as:
\begin{equation}\label{eq:eva}
    \varphi_u = \log(~ \sigma(W_e(\mathbf{\Psi}_u^T \odot \hat{\theta}_u)) ~).
\end{equation}
Obviously, by maximizing the value of $\varphi_u$, $\mathbf{\Psi}_u$ will get aligned with user interest as much as possible.

\subsubsection{Validator}
Inspired by the idea of generative adversarial network, A binary classifier $\mathrm{VAL}$ is trained for validation, which determines whether a vector comes from the real items' representation $\{\theta_v\}_V$ (with label 1), or those synthesized retrieval keys $\{\mathbf{\Psi}_u\}_U$ (with label 0). Once deployed, the validator takes a retrieval key $\mathbf{\Psi}_u$ and calculate its log-likelihood of being positive:
\begin{equation}\label{eq:val}
    \omega_u = \log(~ \sigma(~\mathrm{VAL}(\mathbf{\Psi}_u)~) ~).
\end{equation}
Apparently, $\mathbf{\Psi}_u$ will be within the valid scope of real item's representation when $\Omega_r$ is close to zero, thereby resisting the false reward from evaluator.

One more special thing about the validator is that it needs to be iteratively adapted along with the training progress of Actor, as the retrieval key's distribution is changed from time to time. Particularly, everytime one round of training is completed for the Actor, the validator will be refined with the up-to-date $\{\mathbf{\Psi}_u\}_U$.

\subsubsection{Referencer}
The referencer is a parameter-free function, which measures the similarity (e.g., Cosine) between the retrieval key and the representation of its high-relevance candidate. The range of similarity is mapped to (0, 1) so as to keep consistent in scale with other rewards:
\begin{align}\label{eq:ref}
\begin{split}
& \gamma_u = \log(~ \mathrm{sim}(\mathbf{\Psi}_u, \theta_{\acute{v}}) ~), \\
\text{where }  \mathrm{sim}( & \mathbf{\Psi}_u, \theta_{\acute{v}} ) = 0.5 * (~1+cos(\mathbf{\Psi}_u, \theta_{\acute{v}})~).
\end{split}
\end{align}
Here $\acute{v}$ indicates the user's consumed item in reality, thus being qualified to be a high-relevance candidate. By maximizing the value of $\gamma_u$, the retrieval key and candidate's representation will get close to each other as much as possible w.r.t. the chosen similarity.

\begin{algorithm}[t]
\textbf{(1) Pre-training}: \\ 
\While{\emph{not converge}}{
\For{$u \in U$}
    {
    \For{$v \in V_u$}
        {
            encode $v$ to be $\theta_v$ as Eq. \ref{eq:cnn} and \ref{eq:at-cnn}\;
            encode $u$ to be $\theta_u$ as Eq. \ref{eq:at-user} and \ref{eq:cat-user}\;
            encode negative samples as $\{\theta_{v'}\}$\;
            calculate the overall binary cross-entropy loss as Eq:\ref{eq:rec}: 
            $\mathcal{E}_r \leftarrow \varphi_{v,u} - \sum_{v'}\varphi_{v',u}$\;
            update recommendation model (evaluator, user encoder, item encoder) w.r.t. $\nabla \mathcal{E}_r$\;
        }
    } 
}

\textbf{(2) Validator's Initialization}: \\ 
\While{\emph{not converge}}
{
    \For{$v \in V$}{
        encode $v$ as $\theta_v$ as Eq. \ref{eq:cnn} and \ref{eq:at-cnn}\;
        synthesize the retrieval key $\mathbf{\Psi}_u$ as Eq. \ref{eq:syn} based on a piece of randomly sampled user history\;
        calculate the overall binary cross-entropy loss as Eq. \ref{eq:val}: $\mathcal{E}_v \leftarrow \log(\mathrm{VAL}(\theta_v)) - \log(\mathrm{VAL}(\mathbf{\Psi}_u))$\;
        update validator w.r.t. $\nabla \mathcal{E}_v$\;
    }
}

\textbf{(3) Actor's Training}: \\ 
\While{\emph{not converge}}{
\For{$u \in U$}{
    \For{$v \in V_u$}{
        generate retrieval key $\mathbf{\Psi}_u$ as Eq. \ref{eq:syn}\;
        calculate compound reward $\mathcal{R}$: $\varphi_r + \omega_r + \gamma_r$ as Eq. \ref{eq:eva}, \ref{eq:val}, \ref{eq:ref}.\;
        update Actor w.r.t. $\nabla \mathcal{R}$\;
    }
}
refine validator as step (2)\;
}

\caption{\bf Recallnet's Training Process}
\label{alg:1}
\end{algorithm}

\subsection{Training of Recallnet}
Putting together every component of Actor and Critic, Recallnet's training process is summarized as Algorithm \ref{alg:1}. 

Firstly, the pre-training step is carried out, where our deep recommendation model is learned to capture user interest. Notice that the item encoder, user encoder and evaluator will participate the pre-training process; therefore, they will all be learned from such an operation.

Secondly, the validator is initialized, which is to distinguish the real items representations generated by the item encoder and the retrieval keys generated by the initial synthesizer. 

Thirdly, the Actor is trained: given each user's consumed item, the retrieval key is generated based on her history before the consumption; then the compound reward is produced by the Critic so that the Actor can be updated w.r.t. the its gradient. Since Synthesizer and Item encoder's updates will change the distribution of retrieval key and item representation, the original validator will gradually expire; therefore, it will be adapted iteratively along with Actor's update.

\begin{table}[t]
\centering
\includegraphics[width=0.99\linewidth]{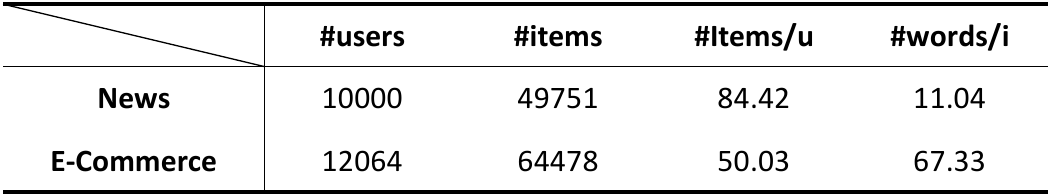}
\caption{Dataset Statistics: total number of users and items, average number of item per user, and average number of word per item.}
\label{tab:stats}
\end{table}

\begin{table*}[t]
\centering
\includegraphics[width=1.0\linewidth]{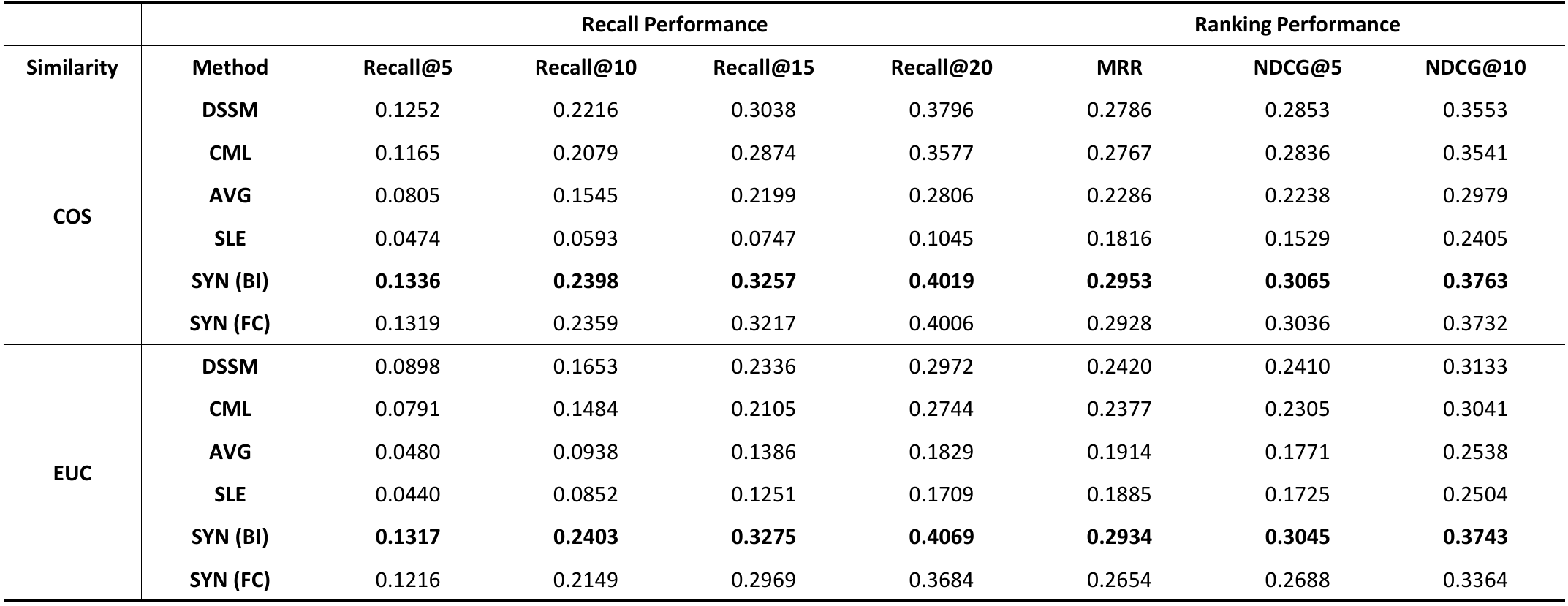}
\caption{Experiment Result On News Dataset (top values marked in bold).}
\label{tab:exp-news}
\end{table*}

\section{Experimental Studies}\label{sec:exp}
\subsection{Experiment Settings}
Experimental studies are carried out for the evaluation of retrieval quality. As a general retrieval pipeline for all the methods in comparison: retrieval key and item representation will be generated in the first place based on both parties' inherent features; then items are sorted according to their representations similarities with the retrieval key; finally, items with the top-K similarity will be selected as the candidates. Detailed configuration about the experiments are introduced as follows.

\subsubsection{Baselines.} Two groups of baselines are compared in our experiments. On the one hand, the \textbf{metric-learning} based methods are taken into comparison, which learn to represent user and item in the common latent space. Particularly, two representative approaches are adopted: one is based on Deep Structured Semantic Model (DSSM) \cite{huang2013learning}, the other one is adapted from Collaborative Metric Learning (CML) \cite{hsieh2017collaborative}. 
\\
$\bullet$ In \textbf{DSSM}, user and item representations are generated by encoding their raw features via two independent networks: the user encoder and item encoder. 
\\
$\bullet$ In \textbf{CML}\footnote{An adaption is made here, as item embedding is used by the original CML. However, the item embedding incurs huge information loss in our experiment, which severely limits its performance.}, user is represented by the embedding vector, which is associated with her ID; while item is still represented with item encoder as DSSM. 
\\
For the sake of fair comparison, user/item encoders in DSSM/CML will use the same structures as our proposed methods, which are illustrated in Section \ref{sec:ins}. 

On the other hand, it is still popular in practice where candidate relevance is directly derived from raw features (e.g., those based on keyword similarity). As a result, we consider those \textbf{learning-free} methods, where user/item's representations are acquired beforehand, instead of specifically learned for candidate retrieval. 
Two representative approaches are adopted: one is based on average word embedding (AVG), the other one uses the sentence level embedding from BERT (SLE). \\
$\bullet$ In \textbf{AVG}, an item is represented as the average vector of its words' embeddings (GloVe-300\footnote{https://nlp.stanford.edu/projects/glove/} is used in our experiment), and a user is represented as the average of her items' embeddings. Despite of simplicity, such a method is a common and effective baseline in many NLP tasks \cite{socher2013reasoning,socher2013recursive,le2014distributed}. \\
$\bullet$ In \textbf{SLE}, an item is represented via the sentence level embedding of BERT \cite{devlin2018bert}, i.e., the embedding of token $\mathrm{[CLS]}$ is used as the item's embedding. Meanwhile, and a user is still represented as the average of her items' embeddings as AVG.\\
Notice that the context feature in recommendation request is not directly comparable with item's text feature, thus they are ignored in AVG and SLE, where user embedding is used as the retrieval key.

\begin{table*}[t]
\centering
\includegraphics[width=1.0\linewidth]{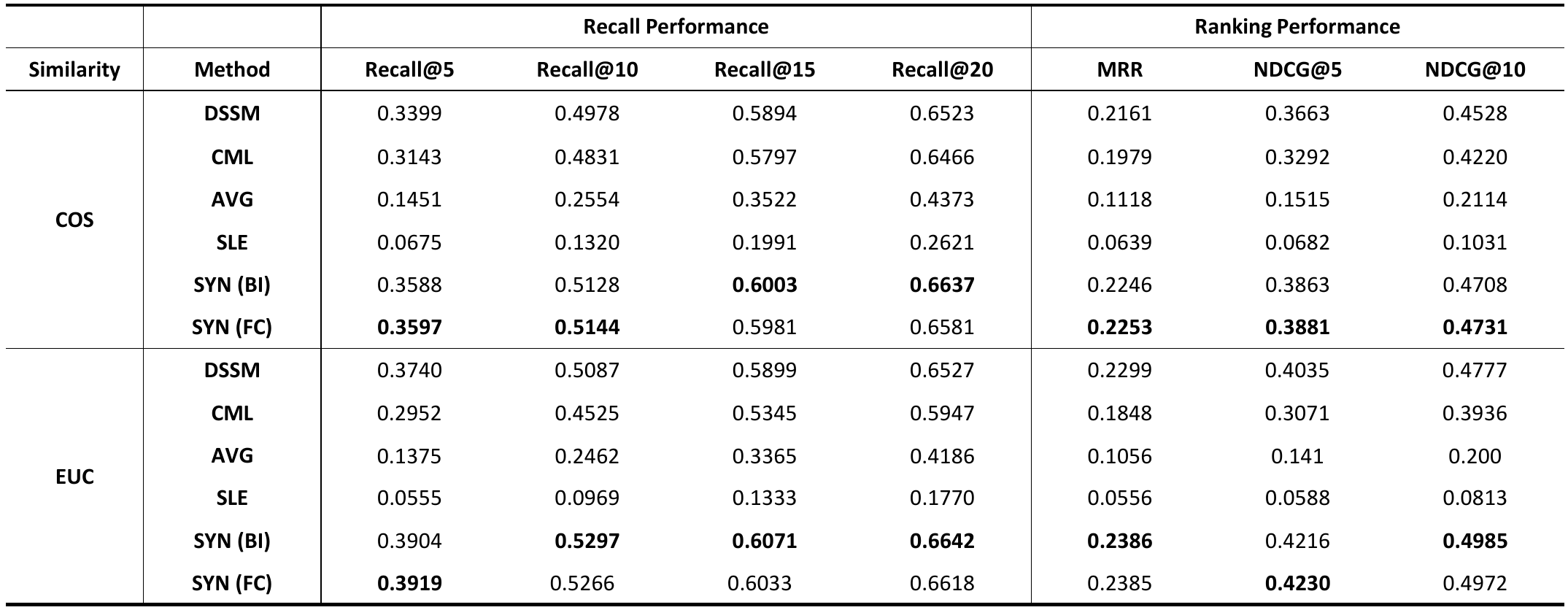}
\caption{Experiment Result On E-Commerce Dataset (top values marked in bold).}
\label{tab:exp-commerce}
\end{table*}

\subsubsection{Variations of Our Approach}
As is introduced, Recallnet can be instantiated in different ways. Therefore, alternative implementations are systematically tested so as to verify its generalizability. 
$\bullet$ \textbf{Variational Similarity Measurement.} Cosine similarity (Eq. \ref{eq:ref}) is chosen as default similarity measurement; besides, Euclidean similarity is also included in the experiments.
\\
$\bullet$ \textbf{Variational Form of Evaluator.} The bi-channel feed-forward network (BI) introduced in Section \ref{subsec:eva} is used as default evaluator; meanwhile, the fully-connected feed forward network (FC) is also taken into account, where user and item representations are concatenated along the column and processed by a 2-layer feed-forward network for final logit. 
\\
$\bullet$ \textbf{Variational Form of User Encoder.} In addition to our default user encoder introduced in Section \ref{sec:ins-user}, the way of user representation in CML is also considered, where users are represented by the embedding vectors associated with their IDs.
\\
$\bullet$ \textbf{Variational Input.} Apart from user history, there can be other available information when recommendation is to be made, such as context and user's intent. The auxiliary information is encoded in parallel with user history, and the encoded vectors of both parts' are concatenated along column for the final user representation.

\subsubsection{Metrics.} The following three evaluation measurements are considered in our experiments. \\
$\bullet$ Recall Performance. The retrieval quality is directly reflected by its recall rate, as the ultimate goal of retrieval is to obtain all the quality candidates, instead of giving the final recommendation list. The recall rate is measured with Recall@K, where K is scale of retrieval set.   \\
$\bullet$ Ranking Performance. To know more about the retrieval precision of different methods, the ranking performance is further compared, which is measured with MRR and NDCG@K. \\

\subsubsection{Datasets.} 
Two real-world datasets are used in our experiments. 
One is the industrial dataset \footnote{https://www.msn.com/en-us/news} from MSN News, which records users' news browsing behaviors. In this dataset, each user is associated with her browsed news articles in history, and each article is associated with its titles; additionally, user's intent is explicitly specified for each of her browsing behavior, i.e, the type of news (e.g., political or financial news) she's looking for. In our experiment, user history is used as by default; and user's intent is used while evaluating the effect of auxiliary information.
Another public dataset of Amazon reviews on Movies and TV\footnote{http://jmcauley.ucsd.edu/data/amazon/} (referred as E-Commerce) is adopted in the experiment, which records users' online shopping behaviors: each user is associated with her purchased items in history, and each item is associated with its title and description. Detailed statistics about both datasets are illustrated as Table \ref{tab:stats}.


\begin{table*}[t]
\centering
\includegraphics[width=1.0\linewidth]{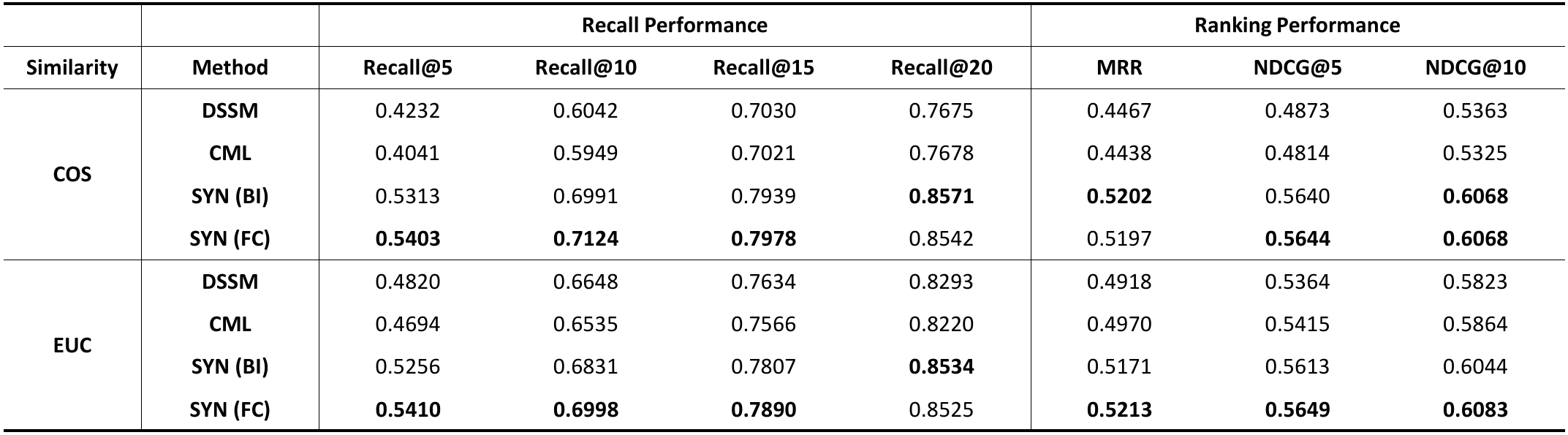}
\caption{Experiment Result Using Auxiliary Input (top values marked in bold).}
\label{tab:exp-aux}
\end{table*}

\begin{table*}[t]
\centering
\includegraphics[width=1.0\linewidth]{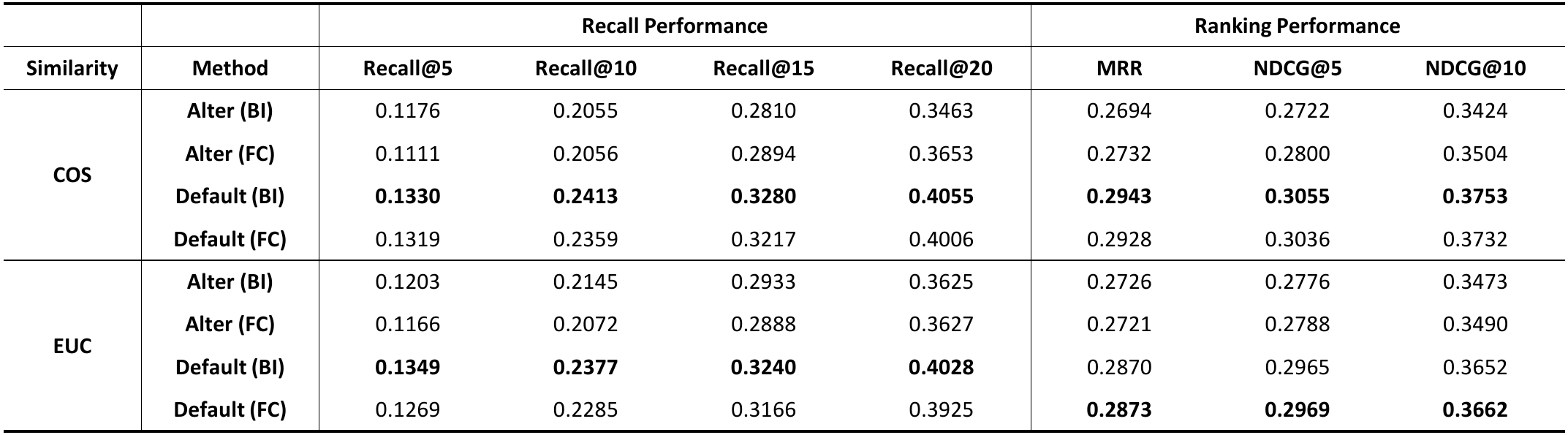}
\caption{Experiment Result Using Different User Encoder; Default: with default user encoder, Alter: with user encoder from CML (top values marked in bold).}
\label{tab:exp-ue}
\end{table*}

\subsection{Experiment Analysis}
\subsubsection{Findings From Main Results}\label{sec:exp-main} 
Main experimental results on both datasets are shown by Table \ref{tab:exp-news} and \ref{tab:exp-commerce}. As is can be observed, variations of Recallnet, SYN (B) and SYN (F) consistently outperform all the baselines in terms of recall performance, indicating that better candidates can be retrieved with our proposed method. It may also be inferred that Recallnet's superiority on recall performance is resulted from its higher capability of identifying quality candidates, as top ranking scores can always be generated from it. Besides, it can be observed that all the learned representations (with SYN (*), DSSM, CML) outperform those learning-free methods significantly. The explanation about this phenomenon is twofold. For one thing, learning-free approaches, like \textbf{AVG}, are too simple to exploit user and item's raw features effectively, thus unable to identify candidate's relevance in fine-granularity. For another thing, although some other learning free methods, like SLE, are sophisticated to fully utilize raw features, their generated user/item representations may not relevant in terms of similarity, thus unsuitable for candidate retrieval task.


Apart from the above obvious observations, some other interesting phenomenons can also be derived from the main result.
\\
$\bullet$ \textbf{Effect of Similarity Measurement.}
For both similarity measurements (COS and EUC), SYN (*) consistently gives rise to the highest recall/ranking performances for all the testing cases; in addition, Recallnet's fluctuations across different similarity measurements are comparably smaller than those of baselines. 
The above phenomenons indicate that Recallnet is robust to the change of similarity measurement. One probable explanation is that Recallnet takes advantage of multiple supervision signals in its training process, thus making it less sensitive to each individual one. A more detailed analysis is to be made in Section \ref{sec:exp-ca}.
\\
$\bullet$ \textbf{Effect of Evaluator's Form.}
For both forms of evaluators (BI and FC), consistent improvements are achieved over the baselines under the same setting, indicating that all forms of evaluators contribute substantially to the Recallnet's performances. However, distinct results might be generated by different forms of evaluators, which suggests that Recallnet's performance can be optimized by selecting more effective evaluators. 

\begin{table*}[t]
\centering
\includegraphics[width=1.0\linewidth]{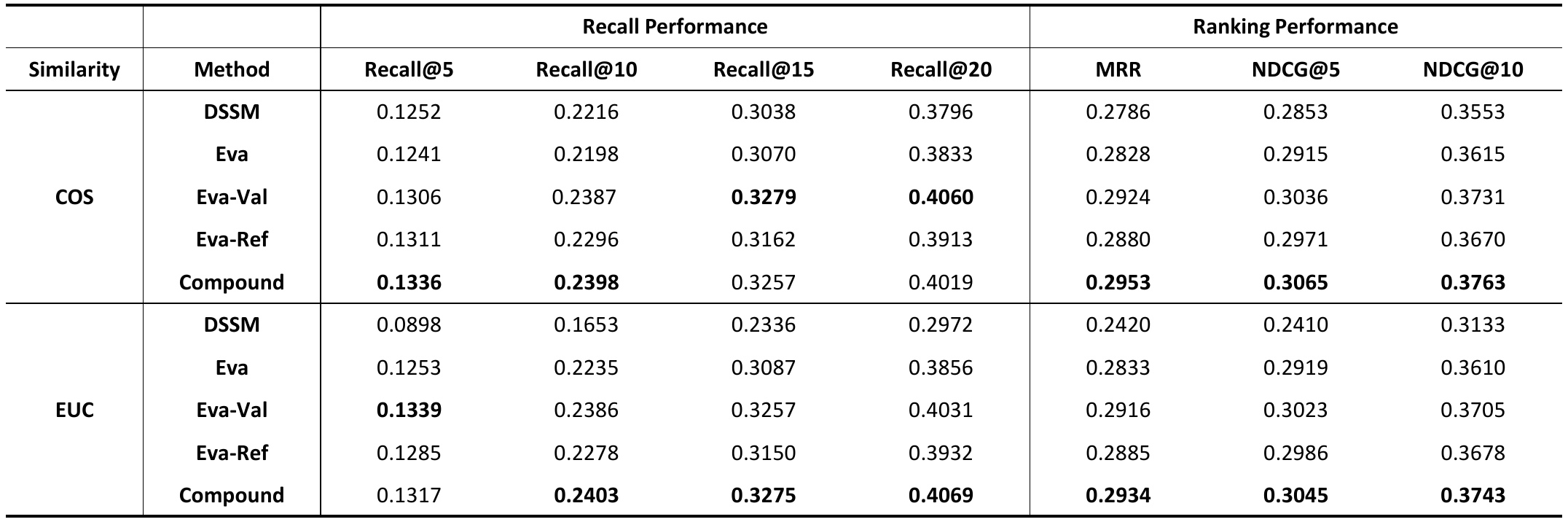}
\caption{Experiment Result For Component Analysis (top values marked in bold).}
\label{tab:exp-ca}
\end{table*}

\subsubsection{Additional Studies}\label{sec:exp-add}
A series of additional studies are carried out for the complement of the main results. Because of redundant observation, analysis is only carried out for the result on News dataset. 
\\
$\bullet$ \textbf{Recallnet with Auxiliary Input.} As introduced before, user's intent is available for the News dataset, and it is adopted as our auxiliary input. The intent is virtually a categorical variable specifying the type of news the user's looking for, therefore it is represented by the embedding vector associated with its ID. According to the experiment results demonstrated in Table \ref{tab:exp-aux}, SYN (*) still gives rise to the best recall/ranking performances in contrast to the baselines (the learning-free methods are omitted due to their incapability of using auxiliary input). Besides, recall/ranking performances are remarkably improved (compared with those in Table \ref{tab:exp-news}) thanks to the presence of additional information. As a result, it validates that Recallnet is able to effectively exploit auxiliary input for better retrieval performance.
\\
$\bullet$ \textbf{Recallnet with Different User Encoder} Performances with different user encoders are demonstrated in Table \ref{tab:exp-ue}. As it can be observed, SYN with our default user encoder consistently outperform those with CML's user encoder. Together with our conclusion in Section \ref{sec:exp-main}, we may have the following conclusion. That as a general candidate retrieval paradigm, Recallnet consistently outperforms those metric-learning baselines under the same settings (similarity measurement, user/item encoders); meanwhile, the performance of Recallnet itself can be further enhanced by selecting more appropriate configurations, such as forms of evaluator and user/item encoders.
\\
$\bullet$ \textbf{Alignment with User Interest.} In addition to our metrics on recall/ranking performance, we would also like to know user's degree of interest towards the retrieved candidates, which can be measured as the negative log-likelihood of the top K retrieved candidates (denoted as LL@K):
\begin{equation}
    \mathrm{NLL@K} = \sum\nolimits_{0\leq j\leq K} -1 * \log(P_{u_i}(v_j)),
\end{equation}
where $P_{u_i}(x_j)$ stands for user's probability of being interested in item $v_i$. Apparently, a smaller NLL@K indicates a larger degree of interest. Since there is no way to acquire user's exact interest, the well-trained evaluator used by SYN (B) is employed for approximation.
\\
According to the demonstrated result in Table \ref{tab:exp-ui}, the degree of interest is almost aligned with the recall/ranking performances reported in Table \ref{tab:exp-news}, despite that the employed user model is not fully accurate. As a result, it indicates that the retrieved candidates from Recallnet better meet user's interest.

\subsubsection{Component Analysis for Critic}\label{sec:exp-ca}
Experiments are conducted to evaluate each individual component's effect in Critic, where EVA, VAL, REF indicate the presence of evaluator, validitor and references in Critic\footnote{The combination of Val-Ref is not considered as validator needs to work along with evaluator.}; while Compound stands for the inclusion of all these components. Because of duplicated observations, results are only reported for SYN with evaluator BI.

As demonstrated in Table \ref{tab:exp-ca}, top performances are achieved by Compound in most testing cases.
Meanwhile, improvements can be observed when evaluator work jointly with either validator or referencer. Therefore, it indicates that both components contribute substantially to Recallnet's performance.
Moreover, Recallnet's performance is already no lower than the best baseline (DSSM) merely with the evaluator; and in some cases, top retrieval results can be obtained merely with evaluator and validator. Both phenomenons suggest that maximizing user's interest towards the retrieval key is crucial for the retrieval quality. In fact, such a point is also consistent with our problem formulation in Eq. \ref{eq:lag}.

\subsubsection{Summarization}
Major findings of the experimental studies are summarized with the following points.

$\bullet$ Consistent and remarkable improvements are achieved by Recallnet in terms of recall/ranking performances, whereby validating its effectiveness on retrieving quality candidates.

$\bullet$ As a general candidate retrieval paradigm, Recallnet can be tuned flexibly for the optimal performance by selecting the most suitable configuration of each specific scenario. 
    
$\bullet$ All the components in Critic contribute substantially to Recallnet's performance, which jointly gives rise to its superior candidate retrieval quality.

    
    

\section{Related Work}\label{sec:rel}
In this section, related studies are reviewed from two perspectives: deep recommendation algorithms and candidate retrieval.

\subsection{Deep Recommendation Algorithms}
Leveraging the recent progress of deep learning, today's recommendation algorithms become more and more proficient in capturing user's underlying interest. Roughly speaking, deep learning techniques contribute to the development of recommendation algorithms in two ways. For one thing, thanks to deep neural networks' superior capability on function approximation, complex user-item relationships, e.g., high-order feature interaction ~\cite{cheng2016wide,guo2018deepfm,lian2018xdeepfm,zhou2018deep,wang2017deep}, temporal behavioral patterns \cite{zhu2017next,zhou2018eolvedeep}, can be effectively learned from user's behavioral data. For another thing, the employment of deep neural networks facilitates the effective exploitation of diverse data, such as textual \cite{Tay2018multi,wu2019neural}, visual \cite{Wang2017image,Joonseok2018collaborative,xu2018hulu} relational \cite{Wang2018billion,Chang2015heterogeneous,Ying2018graph} information, and common-sense knowledge from KB ~\cite{zhang2016collaborative,wang2018dkn}.
It's noticeable that most of these advanced algorithms mainly emphasize the ranking efficacy, yet contribute little to the candidate retrieval due to their inherent limitation on temporal scalability. 

\begin{table}[t]
\centering
\includegraphics[width=0.99\linewidth]{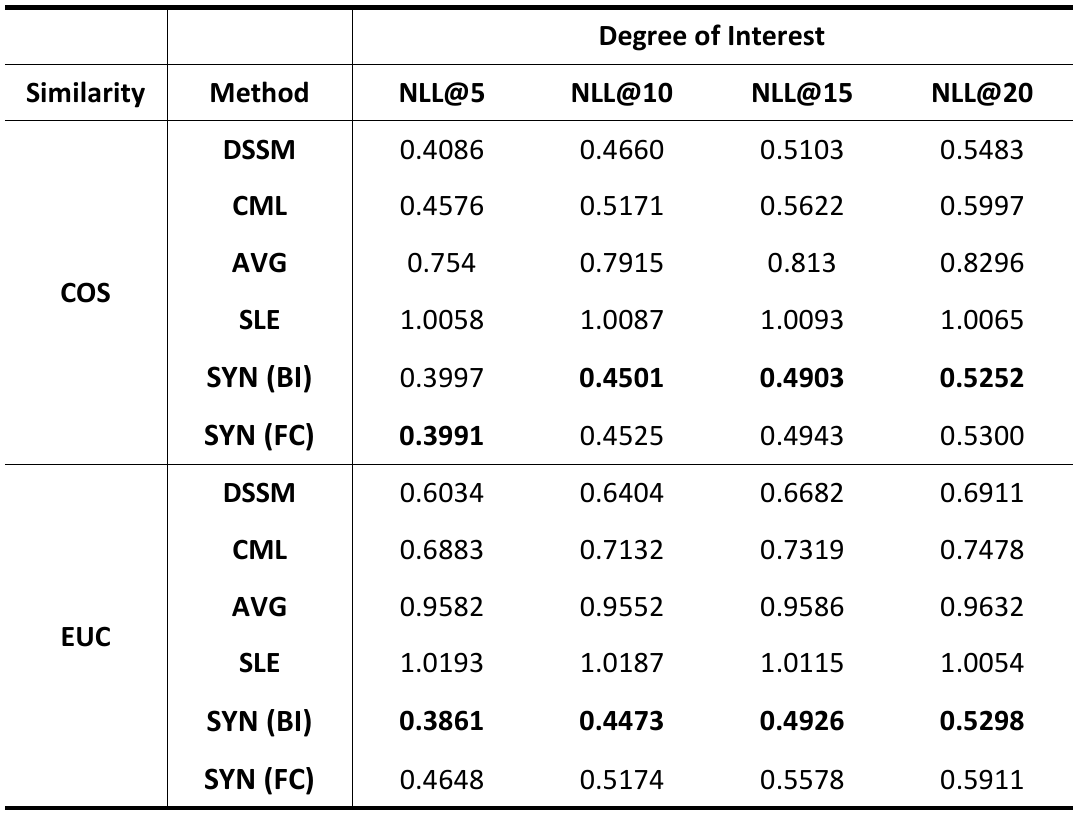}
\caption{Degree of User Interest (top values marked in bold).}
\label{tab:exp-ui}
\end{table}

\subsection{Candidate Retrieval}
\subsubsection{Conventional Way of Candidate Retrieval.} As discussed, candidate has to be selected in realtime from a tremendous pool of items. Therefore, mainstream candidate retrieval approaches would take advantage of structurized data so as to achieve feasible running efficiency. In early days, one of the most well-known representatives is based on inverted-index \cite{buttcher2016information} (still widely applied in practice), where items are indexed w.r.t. a certain type of raw feature (e.g., keywords), and candidates are retrieved for a user if there exist a shared feature value. 
Later on, improved methods are consecutively proposed, where multiple raw features can be jointly utilized for candidate retrieval, and personalized features weights are learned for more precise retrieval \cite{borisyuk2016casmos,anagnostopoulos2008effective,asadi2013effectiveness}. In more generalized setting, user and item's relevance is derived based on their feature similarity. As a result, candidate retrieval can be conducted with even higher flexibility; meanwhile, thanks to the superior index structures like LSH and KNN graph \cite{ChenW18,WangL12, sugawara2016approximately}, the retrieval operation can be efficiently conducted with O(1) time complexity. 


\subsubsection{Metric Learning.} Metric learning is general machine learning paradigm \cite{lebanon2006metric,weinberger2009distance}, which is developed to represent entities such that those from the same class can be mutually close to each other in the representation space. 
Obviously, it turns out to be a natural choice for candidate retrieval,
as user and item can be represented in the common space and their relevance will be measured by representations' similarity \cite{schroff2015facenet,hsieh2017collaborative,Joonseok2018collaborative}.  
In contrast to those conventional methods,
metric learning is able to identify quality candidates more accurately, as representations are carefully learned from user-item interactions, and raw features can be better exploited on top of more advanced structures. 

In fact, both Recallnet and metric learning will represent user as the retrieval key for candidate retrieval. However, there is fundamental distinction on how the retrieval key is generated. For one thing, metric learning merely cares about the overall similarity between the retrieval key and user's consumed items in history, whereas user's interest towards to retrieval key itself is not taken into account.
As a result, the retrieval key may stray away from user's interested region, which will falsely introduce inaccurate candidates and impair the retrieval quality.
On the other hand, Recallnet makes user's interest towards the retrieval key a priority, which is to be optimized simultaneous along with the similarity part.
Therefore, more accurate candidate retrieval can be delivered. In this place, a toy example is presented for better illustration.

\begin{figure}[t]
\centering
\includegraphics[width=0.92\linewidth]{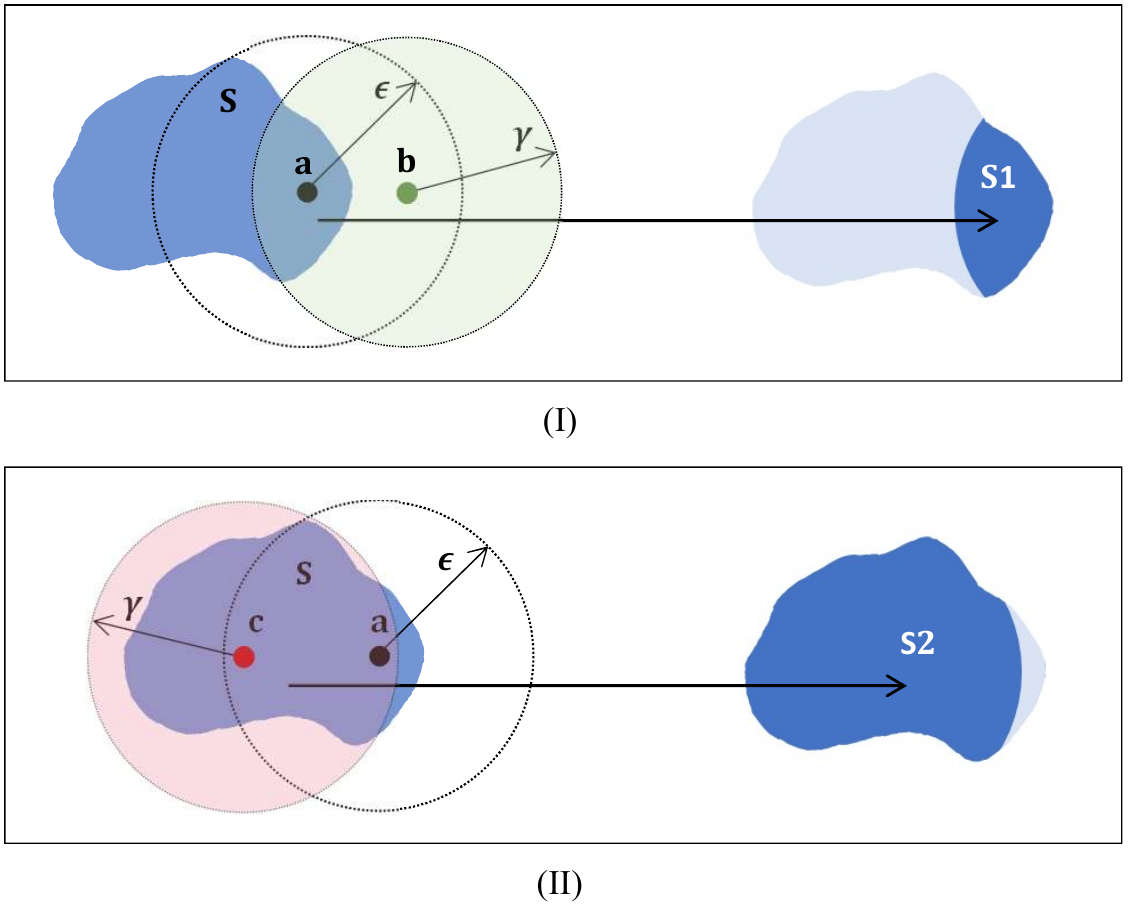}
\caption{Toy example for the comparison of learning to generate retrieval key based on (I) metric learning and (II) Recallnet, respectively (better viewed in colour).}
\label{fig:example}
\end{figure}

\begin{example}
Suppose that representations of a user's interest items are confined within the blue region \textbf{S} in Figure \ref{fig:example}. Meanwhile, vertex \textbf{a} indicates the representation of the user's consumed item in history.
In metric learning, similarity (or distance) becomes the only factor to be considered, and it will determine the whole neighborhood of \textbf{a} (i.e., \textbf{circle} \textbf{(a, $\epsilon$)}) to be the potential region for user representation. Consequently, user's representation could be falsely mapped to vertex \textbf{b}, which is out of user interested region despite its similarity with \textbf{a}; 
and all the items within \textbf{b}'s neighbourhood (i.e., \textbf{circle} \textbf{O(b, $\gamma$)}) will be retrieved as candidates. Apparently, only limited part of user's interested area can be covered (i.e., \textbf{s1}), and many of the relevant candidates could be left out from the retrieval result.
On the other hand, by taking user interest into account, Recallnet will identify vertex \textbf{c} to be a much more appropriate user representation, as it is not only similar with \textbf{a} but also accurately aligned with user interest. Therefore, items within \textbf{circle} \textbf{(c, $\gamma$)} will be selected as the retrieval result, where much more of user's interested items (i.e., those within \textbf{s2}) can be obtained.
\end{example}

\section{Conclusion}\label{sec:con}
In this work, a novel paradigm Recallnet is proposed for personalized candidate retrieval, 
where the virtual item representation (i.e., retrieval key) is synthesized optimally for the efficient acquisition of high-quality candidates. 
With the developed Actor-Critic infrastructure, user's underlying interest becomes accurately aligned with the similarity between retrieval key and item representation. Therefore, high-quality candidates can be effectively identified and efficiently retrieved via similarity search.
Extensive empirical studies are carried out with real-world datasets, whose results validate the effectiveness of our proposed method. 



\bibliographystyle{Format/ACM-Reference-Format}
\bibliography{ref}

\end{document}